\documentclass[twocolumn, 10pt, a4paper]{article}
\usepackage[absolute]{textpos}
\usepackage[top=3cm,bottom=3cm,left=2cm,right=2cm]{geometry}
\usepackage{amsmath,amssymb,mathtools,dsfont,emptypage,graphicx,dcolumn,bm,booktabs,colortbl,collcell,enumerate,float,verbatim,multirow,xparse,slashed,mathrsfs,color}

\usepackage{cite}
\usepackage{tabularx}
\usepackage{multirow}
\usepackage{graphicx}
\usepackage{subcaption}
\usepackage{appendix}
\usepackage{booktabs} 
\usepackage{colortbl}
\usepackage{collcell}
\usepackage{xparse}
\usepackage{relsize}
\usepackage{multicol}
\usepackage{titlesec}
\usepackage{float}
\usepackage[labelfont=bf, labelsep=period]{caption}
\usepackage[compat=1.1.0]{tikz-feynman}
\usepackage{contour}
\usepackage{footmisc}
\usepackage{scrextend}
\deffootnote[1em]{1.5em}{1em}{\textsuperscript{\thefootnotemark}\,}
\usepackage{etoolbox} 
\titleformat{\section}
  {\normalfont\large\bfseries} 
  {\thesection. }{.1em}{}         

\titleformat{\subsection}
  {\normalfont\normalsize\bfseries}
  {\thesubsection. }{.1em}{}

\titleformat{\subsubsection}
  {\normalfont\normalsize\bfseries}
  {\thesubsubsection. }{.1em}{}

\newcommand{\virg}[1]{``#1''}

\def\min{\mathsmaller{\mathrm{min}}}
\def\max{\mathsmaller{\mathrm{max}}}

\def\DM{\mathsmaller{\mathrm{DM}}}
\def\BH{\mathsmaller{\mathrm{BH}}}
\def\LOS{\mathsmaller{\mathrm{LOS}}}
\def\DIS{\mathsmaller{\mathrm{DIS}}}

\def\NR{\mathsmaller{\mathrm{NR}}}

\def\GS{\mathsmaller{\rm GS}}

\definecolor{myred}{cmyk}{0,0.8,0.5,0.5}
\definecolor{myblue}{cmyk}{0.8, 0.4, 0, 0.2}
\definecolor{mygreen}{rgb}{0.27, 0.64, 0.48}
\definecolor{mygray}{gray}{.95}
\definecolor{mygray}{gray}{.95}
\definecolor{niceorange}{rgb}{0.9, 0.3, 0.2}
\definecolor{nicepurple}{rgb}{0.7, 0.0, 0.4}
\definecolor{royalblue}{rgb}{0.0, 0.14, 0.4}
\usepackage[linktoc=page,bookmarks=false,colorlinks=true,linkbordercolor=myblue,citebordercolor=myblue,citecolor=myblue,urlbordercolor=myblue, urlcolor = myblue, anchorcolor=myred,filecolor=myred,linkcolor=myred,menucolor=mygreen]{hyperref}

\begin{document}

\renewcommand*{\thefootnote}{\fnsymbol{footnote}}
\twocolumn[{%
\begin{@twocolumnfalse}
\begin{center}
{\bf\Large Diffuse astrophysical neutrinos from dark matter around blazars} \\
[5mm]
Andrea Giovanni~De Marchi$^{a,b}$ \footnotemark[1], 
Alessandro Granelli$^{a,b}$ \footnotemark[2],
Jacopo Nava$^{a,b}$ \footnotemark[3]  
and
Filippo Sala$^{a,b}$
\footnotemark[4]
\\
$^{a}$\,{\it \small Dipartimento di Fisica e Astronomia, Università di Bologna, via Irnerio 46, 40126 Bologna, Italy; and} \\
$^{b}$\,{\it \small INFN, Sezione di Bologna, viale Berti Pichat 6/2, 40127, Bologna, Italy.}
\end{center}
\begin{center}
    {\bf Abstract}
\end{center}
\noindent
Neutrinos from blazars can originate from inelastic scatterings between protons within their jets and sub-GeV dark matter (DM) around them, explaining IceCube detections of neutrinos from TXS 0506+056 that are otherwise challenging for models of its jet.
In this paper we calculate such DM-induced high-energy neutrinos, from TXS 0506+056 as well as from a stacked blazar sample, in the four cases where DM-quark interactions are mediated by a new massive vector, axial, scalar, and pseudoscalar particle. 
Intriguingly, we find that this mechanism can saturate the diffuse astrophysical neutrino flux observed by IceCube at high energies.
Our mechanism will be tested by additional blazar observations and by various searches for sub-GeV DM. \\
\hrule
\vspace{1em}
  \end{@twocolumnfalse}
}]

\renewcommand*{\thefootnote}{\fnsymbol{footnote}}
\addtocounter{footnote}{0} 
\footnotetext[1]{\href{mailto:andreagiovanni.demarchi@gmail.com}{andreagiovanni.demarchi@unibo.it}}
\footnotetext[2]{\href{mailto:alessandro.granelli@unibo.it}{alessandro.granelli@unibo.it}}
\footnotetext[3]{\href{mailto:jacopo.nava2@unibo.it}{jacopo.nava2@unibo.it}}
\footnotetext[4]{\href{mailto:f.sala@unibo.it}{f.sala@unibo.it}, FS is on leave from LPTHE, CNRS \& Sorbonne Universit\'e, Paris, France}

\newpage

\tableofcontents

\section{Introduction} 

\renewcommand*{\thefootnote}{\arabic{footnote}}
\setcounter{footnote}{0}

Dark matter (DM) provides incontrovertible evidence for the need of new physics beyond the established laws of Nature. However, viable DM constituents span dozens of orders of magnitude in both mass and interaction with known standard model (SM) particles~\cite{Cirelli:2024ssz}.
DM candidates lighter than a proton recently became the object of intense investigation, after the lack of detection (so far) for candidates in other mass ranges. This resulted in the proposal of several novel mechanisms to explain the cosmological origin of sub-GeV DM, as well as to detect its SM interactions.

Moving from fundamental physics to astrophysics, the origin of the high-energy neutrinos observed by IceCube and KM3NeT constitutes today a mystery, see e.g.~\cite{Fang:2022trf,Arguelles:2024ncf}. This is the case both for the diffuse \virg{astrophysical} neutrinos and 
for those detected by single blazars, which are black holes with two powerful jets one of which points towards us.
For example, the widely-adopted one-zone
models of such jets fall short in explaining the diffuse astrophysical neutrinos from stacked blazar emissions~\cite{Murase:2018iyl,Rodrigues:2023vbv}, as well as the neutrino detected from the blazar TXS 0506+056~\cite{Gao:2018mnu, MAGIC:2018sak, Cerruti:2018tmc,Cerruti:2018tmc_erratum,Keivani:2018rnh}.

Could these two problems be solved at once, namely could such high-energy neutrinos be the first evidence of a non-gravitational DM-SM interaction?

\medskip

A positive answer to this question was first provided in~\cite{DeMarchi:2024riu}, where we proved that the dominant flux of neutrinos from single blazars could originate from inelastic scatterings between protons within their (one-zone) jets and sub-GeV DM around their black hole. In particular, the neutrinos detected from TXS 0506+056 could be due to DM-proton interactions that are allowed by all existing constraints.
This mechanism is not only far-reaching, because it connects independent mysteries in different domains of science, but it can be tested by the rich and expanding experimental program to search for sub-GeV DM, offering many handles to tell it apart from different explanations of blazar neutrinos that rely on complicating jet and/or blazar models, like~\cite{Liu:2018utd, Righi:2018xjr, Sahakyan:2018voh, Wang:2018zln, Zhang:2019dob, Xue:2019txw, Zhang:2019htg, KhateeZathul:2024tgu, Wang:2024tsd}.
These results~\cite{DeMarchi:2024riu} motivate the exploration of several further avenues. In this paper we focus on the following ones:
\begin{enumerate}[i)]
\item
It is worth going beyond single blazars and computing the stacked neutrino flux from DM-proton scatterings in the jets of many blazars. Can this flux saturate or exceed IceCube observations of diffuse astrophysical neutrinos? What valuable information comes from their comparison?
\item
In~\cite{DeMarchi:2024riu} we not only restricted ourselves to single blazars, but also to DM-SM interactions mediated by a new massive vector with promising but specific choices of its parameters. How do different choices about the mediator impact our DM explanation of neutrinos (both diffuse and from TXS 0506+056), as well as its experimental tests?
\end{enumerate}

After reviewing the properties of blazar jets and of the DM around them in Sec.~\ref{sec:intro_astro}, in Sec.~\ref{sec:neutrinos} we define the four particle DM models of interest and compute the neutrino fluxes they induce, both from TXS 0506+056 and from a stacked sample of blazars. In Sec.~\ref{sec:tests} we discuss some tests of our findings, in Sec.~\ref{sec:summary} we conclude.

\section{Blazars jets and their dark matter}
\label{sec:intro_astro}

\subsection{Blazars and their jet models}
Active galactic nuclei (AGN) are the most powerful
steady luminous sources in the Universe, powered by supermassive black holes (BHs) in their center. 
They are called \textit{blazars} if they present a pair of back-to-back jets, made of relativistic plasma, and one of them forms a small angle $\theta_\LOS \lesssim 15^\circ-20^\circ$ with our line-of-sight (LOS) \cite{Urry:1995mg, Giommi:2011sn, Giommi:2013mck, Padovani:2017zpf}.

Blazars electromagnetic activity is mainly non-thermal, with a spectral energy distribution (SED) of photons that exhibits two distinct peaks: one in the infrared/X-ray band and the other at $\gamma$-ray frequencies \cite{Padovani:2017zpf}, that can be attributed to the non-thermal radiative emission from charged particles in the jets as they propagate through magnetic fields
and ambient radiation. \footnote{
Based on the spectral features in the optical frequencies, blazars are further subdivided into: Flat Spectrum Radio Quasars (FSRQs), which, besides the flat radio spectrum that gives them their name, exhibit strong and broad optical emission lines; and BL Lacs, the optical spectrum of which shows at most very weak lines, if present at all 
\cite{Giommi:2011sn}. Some blazars are classified as \virg{masquerading} BL Lacs, when their optical spectrum is of the FSQR-type but is hidden by the non-thermal jet emission \cite{Giommi:2013mck} (the well-studied blazar TXS 0506+056 is an example of this kind \cite{Padovani:2019xcv}).
}

The emitting regions, in which the particles in the jets are confined, are typically considered as spherical \textit{blobs} that move at relativistic speed $\beta_B$ along the jet axis. 
A particularly compelling category of blazar jet models is that of the hybrid lepto-hadronic ones, according to which both electrons and protons are present in the blob with ultra-relativistic velocities, and the SED arises from a combination of leptonic and hadronic processes (see, e.g., \cite{Cerruti:2020lfj} for a review). A distinctive prediction of this class of models is the production of 
neutrino fluxes from hadronic cascades initiated when protons collide with photons or other protons in the blazar environment. After IceCube's first $\gtrsim 3\sigma$ spatial association between multi-TeV neutrino events and a blazar, TXS 0506+056 \cite{IceCube:2018dnn, IceCube:2018cha, Padovani:2018acg} \footnote{Two significant associations of neutrino signals and TXS 0506+056 have been reported: one single neutrino event in 2017 in coincidence with a six-month $\gamma$-ray flaring episode \cite{IceCube:2018dnn}, and $\sim 13$ events in 2014/2015 \cite{IceCube:2018cha}. The latter, however, were not accompanied by an enhanced electromagnetic activity of the same blazar \cite{Fermi-LAT:2019hte}. Other tentative associations of different neutrino events with other blazars, albeit with lower significance, have also been reported \cite{Kadler:2016ygj, Paliya:2020mqm, Rodrigues:2020fbu, Oikonomou:2021akf, Liao:2022csg, Sahakyan:2022nbz, Fermi-LAT:2019hte, Jiang:2024nwa, Ji:2024dgn, Ji:2024zbv} (see also \cite{Giommi:2021bar, Boettcher:2022dci}). Also, the active galaxy NGC 1068 has shown evidence of neutrino emission
    at 4.2$\sigma$ \cite{IceCube:2022der}, but this object is not a blazar.
    Furthermore, the study~\cite{IceCube:2019cia} suggests a correlation at the level of 3.3$\sigma$ of
    IceCube neutrinos with a catalog composed of 110 known $\gamma$-ray emitters, of which TXS 0506+056 and NGC 1068 are two main contributors.}, 
an intense campaign to explain both the neutrino emissions and the electromagnetic activities of TXS 0506+056 with lepto-hadronic modelling has been carried out \cite{Gao:2018mnu, MAGIC:2018sak, Cerruti:2018tmc,Cerruti:2018tmc_erratum, Keivani:2018rnh} (see also \cite{ Banik:2019jlm, Banik:2019twt, Cao:2019fnn, Das:2022nyp, MAGIC:2022gyf}). 
   The same class of models has been applied to several other blazars, as, e.g., 
    in \cite{Rodrigues:2023vbv}, advancing the idea that the lepto-hadronic framework may constitute a comprehensive description of blazar jet emission.
    
    In the simplest one-zone lepto-hadronic scenarios, all the photon and neutrino production is attributed solely to processes internal to the blobs, without contributions from external structures or material. Although these one-zone models can satisfactorily explain the SED of most blazars, they struggle to reproduce the reported neutrino events from TXS 0506+056 if one aims to preserve the observed electromagnetic activity \cite{Gao:2018mnu, MAGIC:2018sak, Cerruti:2018tmc,Cerruti:2018tmc_erratum, Keivani:2018rnh} (see also \cite{Murase:2018iyl, Reimer:2018vvw, Rodrigues:2018tku, Petropoulou:2019zqp}). In particular, one-zone lepto-hadronic fits to the observed SED of the 2017 TXS 0506+056 flare predict a neutrino flux so low that accounting for the single IceCube event \cite{IceCube:2018dnn} appears as a stroke of luck \cite{Strotjohann:2018ufz}.

    \medskip
    
 In this work, we build upon the idea that DM can have an impact on this picture, as first pointed out in \cite{DeMarchi:2024riu}. A substantial amount of DM is expected to gravitate around the supermassive BHs inside blazars. Should an interaction between DM and protons occur, the collisions of protons with DM particles would lead to proton disintegration, triggering hadronic cascades and neutrino emission substantially larger than that predicted by one-zone lepto-hadronic models, thereby helping to overcome the aforementioned challenge. 
 In this paper, we analyse this DM hypothesis on top of one-zone lepto-hadronic models.

 To compute the flux of neutrinos produced upon DM-proton interaction in blazars, we need to model the energy spectrum of the protons $p$  within the jets. In the context of one-zone
lepto-hadronic scenarios, this is given in the blob frame as $d\Gamma'_p/(d\gamma'_p \, d\Omega')= \kappa_p/(4\pi) {\gamma'}_p^{-\alpha_p} e^{-\gamma'_p/\gamma'_{\max_p}}$, 
being $d\Gamma_p'$ the infinitesimal rate of protons ejected in the blobs along the direction $d\Omega'$ and with $\gamma_p'$ in the range $[\gamma'_p, \;\gamma'_p + d\gamma'_p]$; $\alpha_p\geq 0$ is the slope of the power-law; $\kappa_p$ is an overall normalization constant. We have also included an exponential cut-off at $\gamma'_{\max_p}$.
 Hereafter, primed (unprimed) symbols refer to quantities computed in the blob’s (observer’s) rest frame. 
The spectrum in the blob framde $d\Gamma'_p/(d\gamma'_p \, d\Omega')$ is related to that in the observer's frame by a Lorentz transformation of boost factor $\Gamma_B \equiv (1-\beta_B^2)^{-1/2}$ along the jet axis, which gives  \cite{Wang:2021jic, Granelli:2022ysi} (see also \cite{Gorchtein:2010xa}):
 \begin{equation}
 \label{eq:CRSpectrum}
     \frac{d\Gamma_p}{d\gamma_pd\Omega} 
     = \frac{\kappa_p}{4\pi}\,\gamma_p^{-\alpha_p} \frac{\beta_p(1-\beta_p\beta_B  \mu)^{-\alpha_p} \Gamma_B^{-\alpha_p} \, e^{-\gamma_p/\gamma_p^\max}}{\sqrt{(1-\beta_p \beta_B \mu)^2 - (1-\beta_p^2)(1-\beta_B^2)}}\,,
 \end{equation}
 where $\beta_p =[1-1/\gamma_p^2]^{1/2}$ is the proton's speed and $\mu$ is the cosine of the angle between its direction of motion and the jet axis.

For our analysis, we consider the blazar TXS 0506+056 and a sample of blazars in steady activity. For TXS 0506+056, we adopt the proton spectrum parameters from the one-zone lepto-hadronic fit to its 2017 six-month flare presented in~\cite{Keivani:2018rnh}. For the blazar sample, we use the lepto-hadronic fits from~\cite{Rodrigues:2023vbv}, which model their average steady-state emission.
The relevant jet parameters given in \cite{Keivani:2018rnh, Rodrigues:2023vbv} are the minimal and maximal Lorentz boost factors of the protons $\gamma'_{\min_p}$, $\gamma'_{\max_p}$; the blob Lorentz factor $\Gamma_B$; the LOS angle of the jet $\theta_\LOS$ assumed to be $\theta_\LOS=1/\Gamma_B$; \footnote{In \cite{Keivani:2018rnh}, the LOS angle was taken to be zero, given the assumed relation $\Gamma_B = \mathcal{D}/2$, where $\mathcal{D} \equiv [\Gamma_{\rm B}\left(1-\beta_{\rm B}\cos\theta_{\LOS}\right)]^{-1}$ is the Doppler factor determined by the fit. Here, we prefer to consider a non-vanishing angle and use the condition $\Gamma_B = \mathcal{D}$, which implies $\theta_{\LOS} = 1/\Gamma_B$. Under this assumption, the proton luminosity given in \cite{Keivani:2018rnh} should be multiplied by a factor of 4 for consistency. However, for the sake of conservativeness, we refrain from applying this correction.}
the proton luminosity $L_p= \kappa_p m_p \Gamma_B^2 \int_{\gamma'_{\min_p}}^{\gamma'_{\max_p}} dx\,x^{1- \alpha_p}\,$.
We summarise them in Table~\ref{tab:AGNparameters}. Also shown are the redshift $z$, the luminosity distance $d_L$, the black hole mass $M_{\BH}$, and the Schwarzschild radius $R_S$.

\setlength{\arrayrulewidth}{.4pt}

\setlength{\tabcolsep}{0pt}
\newcolumntype{C}{@{}>{\centering\arraybackslash}X}
\begin{table}
\centering
\begin{tabularx}{\linewidth}{@{}C||C|C@{}}
    \hline
    \multicolumn{3}{|c|}{\rule{0pt}{2ex} \cellcolor{gray!20}\bf Lepto-Hadronic Model Parameters}\\
    \hline
    \rule{0pt}{2.5ex}
{Parameter (units)} & { TXS 0506+056 \cite{Keivani:2018rnh}} & {Sample of blazars   \cite{Rodrigues:2023vbv}}\\ 
\hline
    \rule{0pt}{2.5ex}
$z$ & 0.337 & $\left[ 0.04, 3.41 \right]$\\
$d_L$ (Mpc) & 1765.29 & $\left[ 170.8, 3\times 10^4 \right]$\\
$M_\BH (M_\odot)$  & $3 \times 10^8$ & $\left[ 10^8, 10^9 \right]$\\
$R_S$ (pc) & $3\times 10^{-5}$  & $[10^{-5}, 10^{-4}]$\\
$\Gamma_B$ & 24.2 & $\left[ 3.4, 31.5 \right]$\\
${\theta_\LOS}(^{\circ})$ & 2.37 & $\left[ 1.81, 16.95 \right]$\\
$\alpha_p$ & 2 & 1\\
$\gamma'_{\min_p}$ & 1 & 100\\
$\gamma'_{\max_p}$ & $1.6 \times 10^{7}$ & $\left[ 10^6, 3.1 \times 10^7 \right]$\\
$L_p$ (erg/s)  & $1.85 \times 10^{50}$ & $\left[  10^{44}, 3 \times 10^{50}\right]$\\
$\kappa_p\,(\text{s}^{-1}\text{sr}^{-1})$ & $1.27 \times 10^{49}$ &  $\left[3 \times 10^{37}, 10^{46}\right]$\\
\hline
\end{tabularx}
\caption{The relevant  jet parameters from the lepto-hadronic fits~\cite{Keivani:2018rnh} for the blazar TXS 0506+056 
and the sample of blazars~\cite{Rodrigues:2023vbv} used in our analysis. Values in brackets denote the ranges for the sources in \cite{Rodrigues:2023vbv}.}\label{tab:AGNparameters}
\end{table}

\subsection{Dark matter around blazars}
\label{sec:LOS}
 DM is expected to cluster into spikes in the vicinity of supermassive BHs, like those that power blazars. As demonstrated in \cite{Gondolo:1999ef}, the adiabatic growth of BH at the centre of a spherical DM halo with a power-law density $\rho^\text{halo}_{\DM}(r; \gamma)=\mathcal{N} r^{-\gamma}$ reshapes the halo profile into a steeper distribution
$\rho^\text{spike}_{\DM}(r) = \mathcal{N} R_\text{sp}^{-\gamma} (R_\text{sp}/r)^{\alpha_{\GS}},$ with $\alpha_{\GS}=(9-2\gamma)(4-\gamma)$,  $\mathcal{N}$ a normalisation constant,  $\epsilon(\gamma)\approx 0.1$ for $0.5\leq \gamma \leq 1.5$ \cite{Merritt:2003qc}, and $R_{\text{sp}}\approx \epsilon(\gamma) (M_{\BH}/\mathcal{N})^{1/(3-\gamma)}$ the radial extension of the spike \cite{Gondolo:1999ef}.

We model the total DM profile $\rho_{\DM}(r)$ for the blazars in our study as $\rho_{\DM}(r \geq R_\text{sp}) = \rho_{\DM}^\text{halo}(r)$ with $\gamma = 1$, as for a Navarro-Frenk-White (NFW) distribution \cite{Navarro:1995iw, Navarro:1996gj},
whereas $\rho_{\DM}(2R_S \leq r < R_\text{sp}) = g(r) \rho_{\DM}^\text{spike}(r)$ with $\alpha_{\GS} = 7/3$, and $g(r)=(1-2R_S/r)^{3/2}$ accounting for the capture of DM onto the BH~\cite{Gondolo:1999ef}, upon including relativistic effects  \cite{Sadeghian:2013laa}.

Due to the blazar jet emission outshining the dynamics of the host galaxy, information on the DM distribution around blazars is limited. Consequently, the normalization of the DM profile remains somewhat arbitrary.
Following the same procedure as in \cite{DeMarchi:2024riu}, we fix $R_\text{sp} = R_\star$,  $R_\star\approx 10^6 R_S$ being the typical radius of influence of a BH on stars \cite{Kormendy:2013dxa}.
This normalisation yields, for $\gamma = 1$, $\mathcal{N} \simeq 10^{-6} (M_\odot/R_S^2) (M_\BH/10^8 M_\odot)$.

The total  DM mass contained within $R_{\text{sp}}$ then is $M_{\DM}^{\text{spike}}\simeq 18.5\%\,M_{\BH}$, compatible with typical BH mass estimates \cite{Labita:2006jg, Pei:2021qay}.
Furthermore, our normalisation is more conservative than that adopted in similar studies on DM around AGN~\cite{Gorchtein:2010xa, Wang:2021jic, Granelli:2022ysi, Cline:2022qld, Ferrer:2022kei, Bhowmick:2022zkj, Cline:2023tkp, Herrera:2023nww, Wang:2025ztb}.

All the relevant information on the DM distribution reduces to the following LOS integral \cite{Wang:2021jic, Granelli:2022ysi} (see also \cite{Gorchtein:2010xa}):
\begin{equation}\label{eq:DMColumnDensity}
    \Sigma_{\DM}^\text{spike} (r)\equiv \int_{R_{\min}}^{r} \rho_{\DM}(r') dr',
\end{equation}
where $R_{\min}$ is the minimal radial extension of the jet. 

 Various effects can alter the formation and evolution of the DM spike. For instance, DM annihilations over the BH lifetime $t_{\BH}$~\cite{Gondolo:1999ef} flatten the DM profile at inner radii reducing the slope of the spike to $\leq 0.5$ for $r<r_\text{ann}$~\cite{Shapiro:2016ypb}, $r_\text{ann}$ being defined by the condition $ \rho_{\DM}(r_\text{ann}) = \rho_\text{core}\equiv m_\DM/(\langle\sigma_\text{ann} v_\text{rel}\rangle t_{\BH})$, 

where $\langle\sigma_\text{ann} v_\text{rel}\rangle$ is the DM averaged annihilation cross section times relative velocity.
Other effects can influence the DM spike, such as merging of galaxies~\cite{Ullio:2001fb, Merritt:2002vj} and
the gravity of stars close to the BH~\cite{Gnedin:2003rj, Bertone:2005hw}, and they can even make the DM profile cuspier than the one we employed~\cite{Gnedin:2011uj,DiCintio:2014xia}.

Sizeable uncertainties also reside in $R_{\min}$ and they have a substantial impact on $\Sigma_{\DM}^{\text{spike}}$. 
To effectively account for 
all these 
astrophysical effects on the DM spike, instead of considering different scenarios for each of them, we find it more practical to consider different benchmark cases (BMCs) for $R_{\min}$.
Following \cite{DeMarchi:2024riu}, we set
$R_{\min} = 10^2 R_S$ (BMCI) and $R_{\min} = 10^4 R_S$ (BMCII), based on blazar studies indicating where the jet is likely well-accelerated \cite{Rodrigues:2023vbv}.
We find $\Sigma_{\DM}^\text{spike} \simeq 6.9\times 10^{28} \,[1.5\times 10^{26}] \, (3\times 10^8 M_\odot)/M_\BH \, \text{GeV}\,\text{cm}^{-2}$ for BMC I [BMC II]. 
We calculate that $r_\text{ann} < r_\text{min}$ admits $\left\langle \sigma_\text{ann} v_\text{rel}\right\rangle\lesssim 1.4\times 10^{-26}\,(3.1\times 10^{-31})\,\text{cm}^3\,\text{s}^{-1}(m_\DM/\text{GeV})$, for $t_{\BH} = 10^{10}\,\text{yr}$.

\section{Neutrinos from dark matter-proton interactions in blazars}
\label{sec:neutrinos}

\subsection{Dark matter-proton interactions}
\label{sec:toymodels}

Protons in blazar jets can undergo deep inelastic scattering (DIS) with the surrounding DM particles and break apart. This initiates a cascade of secondary particles that include neutrinos, possibly detectable by IceCube. To compute these neutrino fluxes we first need to specify the DM interactions with the proton's constituents.

We consider DM as a SM singlet Dirac fermion $\chi$ with mass $m_\DM$, interacting with quarks $q$ via a new mediator $Y$ with mass $m_Y$. We assume four possibilities for $Y$, namely a scalar $\phi$ or a pseudoscalar (or axion-like particle, ALP) $a$ or a vector $V$ or an axial-vector $V'$, described by the following Lagrangians 
\begin{eqnarray}
     && \mathcal{L}_{\chi q \phi}  =
      g_{\chi \phi} \, \bar{\chi} \chi \phi + g_{q \phi} \, \bar{q} q \phi,\label{eq:scalar}\\
     && \mathcal{L}_{\chi q a}  =
      i g_{\chi a} \, \bar{\chi} \gamma^5 \chi a + i g_{q a} \, \bar{q} \gamma^5 q a,\label{eq:pseudoscalar}\\
     && \mathcal{L}_{\chi q V}  = 
         g_{\chi V} \, \bar{\chi} \gamma^\mu \chi V_\mu + g_{q V} \, \bar{q} \gamma^\mu q V_\mu,\label{eq:vector}\\
     && \mathcal{L}_{\chi q V'} =
          g_{\chi V'} \, \bar{\chi} \gamma^\mu \gamma^5 \chi V'_\mu + g_{q V'} \, \bar{q} \gamma^\mu \gamma^5 q V'_\mu,\label{eq:axial}
\end{eqnarray}
where, for definiteness, in what follows we set all quark couplings to zero except $g_{uY} = g_{dY} \equiv g_{qY}$.

\subsection{Computation of the neutrino fluxes}
The first computation of the neutrino flux from DM-proton interactions in blazar jets was carried out in~\cite{DeMarchi:2024riu}. 
We review here this mechanism. The single-flavour neutrino flux exiting the blazar as a result of DM-proton DIS can be computed as
\begin{equation}
\frac{d\Phi_\nu}{d\bar{E}_\nu} \simeq 
\frac{1}{3}\frac{\Sigma^\text{spike}_{\DM}}{m_\DM d^2_L}
\int^{\gamma_{\max_p}}_{\gamma_p^\min(\bar{E}_\nu,\, m_\DM)} \!\!\!\!\!\!\!\!\!\!\!\!\!\! d\gamma_p \, \sigma^{\DIS}_{Y \chi p} \, \frac{d\Gamma_p}{d\gamma_p d\Omega}
\! \biggr\rvert_{\theta_{\LOS}}
\!\!\left<\frac{dN_\nu}{d\bar{E}_\nu}\right>
 ,
 \label{eq:neutrino_flux}
\end{equation}
where $\gamma_p^\min(\bar{E}_\nu, m_\DM)$ is the minimum proton boost factor 
necessary to produce a neutrino of energy
$\bar{E}_\nu$, $\gamma_{\max_p}$ is the maximal boost factor of the protons in the observer's frame of reference, related to the same quantity in the blob's frame of reference reported in Table~\ref{tab:AGNparameters},
$\gamma'_{\max_p}$, via a Lorentz boost $\gamma_{\max_p} = \Gamma_\mathrm{B}\gamma'_{\max_p}$, $\sigma^{\DIS}_{Y\chi p}$ is the integrated DM-proton DIS cross section, for mediator $Y = \phi,\,a,\,V,\,V'$, and the overall $1/3$ factor accounts for neutrino oscillations over astronomical distances.
The quantity $dN_\nu/d\bar{E}_\nu$ is the total (i.e. summed over flavours) number of neutrinos per unit of energy produced by each DM-proton DIS, and $\left\langle \cdot \right\rangle$ denotes the average over all possible scatterings with the quarks inside the proton, each weighted by its respective differential cross section. The averaged number of neutrinos is formally defined as 
\begin{equation}
\left< \frac{dN_\nu}{d\bar{E}_\nu}\right> 
= \frac{1}{\sigma^{\DIS}_{Y\chi p}}
\int_{x_\min}^{1} \!\!\!\! dx \, \int_{Q^2_{\mathrm{min}}}^{x\cdot s} \!\!\! dQ^2 \,  \frac{d^2\sigma^\mathrm{DIS}_{Y\chi p}}{dQ^2 dx}
\, \frac{dN_\nu}{d\bar{E}_\nu}
\label{eq:dNdE}
\end{equation}
and depends on $\gamma_p$, $\bar{E}_\nu$ and $m_\DM$ (the masses of the neutrinos can be safely ignored).

We do not compute the integral above analytically but rather we implement each Lagrangian of the considered toy models in \textsc{FeynRules}~\cite{Alloul:2013bka}, simulate the quark-level scattering process $\chi + p \rightarrow \chi + \textrm{jet}$ via \textsc{Madgraph5}~\cite{Alwall:2014hca}, as well as the hadronisation and showering of the outgoing quarks with \textsc{Pythia8}~\cite{Bierlich:2022pfr}. We set all of \textsc{Madgraph5}'s kinematic cuts to zero, as we are interested in all events that produce neutrinos, regardless of their kinematics. Finally, we select the outgoing neutrinos and bin them in energy.

For our computation we used the \virg{CT10}~\cite{Lai:2010vv} PDFs set, which extends down to $x^{\mathrm{pdf}}_\min = 10^{-8}$. The extrema of integration are set by \textsc{Madgraph5}'s aforementioned $Q_\mathrm{min}^2$, and by kinematics, as the parton-level centre-of-mass energy squared $\hat{s} = x\cdot s$ must satisfy $\hat{s} \geq Q^2_\min$ for the scattering to be kinematically allowed, thus setting a lower extremum on $x$, $x^{\mathrm{kin}}_\min$. 
In order to perform the computation, the PDFs must be well defined, hence the overall extremum is $x_\min \equiv \mathrm{max}\{x^{\mathrm{kin}}_\min, x^{\mathrm{pdf}}_\min\}$.

We perform the parton-level simulation and subsequent hadronization and decay in the  $\chi-p$ centre-of-mass frame,
as opposed to the observer's one, thus the following Lorentz boost is needed to go back and forth between the two:

\begin{equation}
    \gamma_\mathrm{COM} = \frac{m_\DM + E_p}{\sqrt{s}}\,.
\end{equation}

Let us now define $\hat{z}$ as the proton direction. Given a final state neutrino $\nu$, the $\hat{z}$-component of its momentum, energy and angle with respect to $\hat{z}$ transform as
\begin{equation}
\begin{aligned}
 p_{\nu, z} & = \gamma_\mathrm{COM}(p^*_{\nu, z}+\beta \bar{E}_{\nu}^*), \\ 
 \bar{E}_\nu &= \gamma_\mathrm{COM}(\bar{E}_\nu^* + \beta p_{\nu, z}^*),\\
 \cos\theta &= \frac{p_{\nu, z}}{\sqrt{{p^*_{\perp}}^2+p_{\nu, z}^2   }},
\end{aligned}
\end{equation}
where (un)starred quantities refer to the centre-of-mass (observer's) frame, $\beta = \sqrt{1-1/\gamma_\mathrm{COM}^2}$, $p^*_{\perp}$ is the component of the neutrino's 3-momentum
in the plane perpendicular to the $\hat{z}$ axis. 
Given the large value of $\gamma_\mathrm{COM}$, $p_{\nu, z}$ will in general be very large with respect to the transversal components of the momentum, thus $\cos\theta$ will be very close to 1, i.e.~for sufficiently high proton energy the outgoing neutrinos are almost collinear with the initial state proton.
For this reason Eq.~\eqref{eq:neutrino_flux} lacks an angular integral, as the only neutrinos that reach the detector are the ones coming from protons closely aligned to the LOS. To formalize this statement, we only consider protons moving within a narrow cone around the LOS direction, with angular opening $\delta$ such that $\cos{\delta} = 1-10^{-5}$, and only select neutrinos whose angle $\theta$ with respect to the proton direction satisfies $\theta \leq \delta$. Within this narrow cone, the jet's proton spectrum is practically constant in angle and can thus be taken out of the angular integral, which reduces to counting all neutrinos that satisfy $\theta \leq \delta$. 

The procedure described so far allows to obtain $\langle dN_\nu/d\bar{E}_\nu\rangle$ of Eq.~(\ref{eq:dNdE}) and the neutrino flux of Eq.~\eqref{eq:neutrino_flux} (where $\gamma_p^\min(\bar{E}_\nu, m_\DM)$ does not need to be computed explicitly).
Our calculation actually underestimates $d\Phi_\nu/d\bar{E}_\nu$, because
i) \textsc{Madgraph5} avoids any $Q^2 \rightarrow 0$ divergent behaviour by only generating events with $Q^2 > Q^2_\mathrm{min} = 4\;\mathrm{GeV}^2$, ii) we did not include the effect of resonances, of particular relevance for $Q^2 \sim$~GeV$^2$, iii) nor the one of the next-to-leading-order processes with sea-gluons in the initial state.

Finally, cosmological redshift modifies the neutrino flux at Earth into
\begin{equation}
    \frac{d\Phi_\nu}{dE_\nu} = (1+z) \frac{d\Phi}{d\bar{E}_\nu}\Bigg|_{\bar{E}_\nu = E_\nu(1+z)},
\end{equation}
where $E_\nu \equiv \bar{E}_\nu/(1+z)$ is the redshifted neutrino energy, i.e. the one that we measure on Earth.

\subsection{Neutrinos from TXS 0506+056}
\label{sec:TXS} 

\begin{figure}[t]
    \centering
    \includegraphics[width=0.46\textwidth]{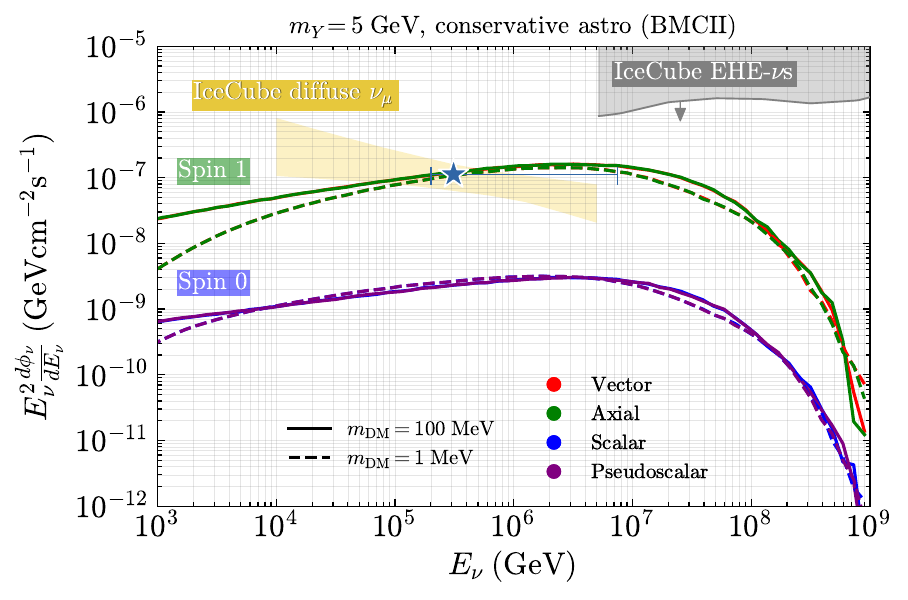}
    \caption{
    Neutrino fluxes from blazar TXS 0506+056 induced by deep inelastic scatterings between protons within its jet and dark matter around it, using as input the jet parameters as fitted in~\cite{Keivani:2018rnh}. The four coloured lines correspond to the four  mediators (vector, axial, scalar and pseudoscalar) of DM-quark interactions defined in Sec.~\ref{sec:toymodels}, for a common choice of coupling $g_{\chi Y}g_{q Y} = 3.5\times 10^{-2} \;(6.5\times 10^{-3})$, DM mass $m_\DM=100$ MeV (1 MeV), mediator mass $m_Y=5$ GeV (where $Y=V,V',\phi,a$) and line-of-sight integral $\Sigma_{\text{DM}} = 1.5 \times 10^{26} \;\mathrm{GeV}\;\mathrm{cm}^{-2}$ (i.e. our conservative benchmark BMCII defined in Sec.~\ref{sec:LOS}). The yellow band is the diffuse astrophysical $\nu_\mu + \bar{\nu}_\mu$ flux measured by IceCube (95\% C.L.) \cite{Abbasi:2021qfz}, the blue star is the best-fit flux from the TXS 0506+056 neutrino and the blue line its uncertainty~\cite{IceCube:2018cha}. In grey the limits set by 9 year of IceCube data from the null detection of extremely-high-energy (EHE) neutrinos~\cite{IceCube:2018fhm}, rescaled to 6 months.
    \label{fig:flux_comparison}
    }
\end{figure}

Having outlined the procedure to compute the neutrino flux
for any model of particle DM and of blazar jets, we now present our results for the blazar TXS 0506+056. We expand on our~\cite{DeMarchi:2024riu} by considering more possible mediators of DM-quark interactions as well as several values of their mass.
In Fig.~\ref{fig:flux_comparison} we show the resulting flux for the four different toy models described in Sec.~\ref{sec:toymodels}, for a common choice of DM parameters as indicated in the figure.

The dependence of $d\Phi_\nu/dE_\nu$ on $m_\DM$ is encoded in $s = 2 m_\DM E_p$. Once $\sqrt{s} >$ few GeV, so that $\nu$ production via DIS is possible, 
$\langle dN_\nu/dE_\nu\rangle$ is determined by $E_p$ and not by $m_\DM$ (we find that it is peaked roughly at $E_p/10$, see~\cite{DeMarchi:2024riu}). Therefore, the high-$E_\nu$ cutoff of our spectra depends on the properties of the jet proton spectrum and not on $m_\DM$.
$\sigma^{\DIS}_{Y \chi p}$ becomes instead suppressed once $\sqrt{s} <$ few tens of GeV, explaining the dependence of $d\Phi_\nu/dE_\nu$ on $m_\DM$ that we find at small $E_\nu$. 
The $\mathcal{O}(50)$ difference in the fluxes, between the spin-0 and spin-1 cases, originates from the $Q^2$-dependences of the associated $\sigma^{\DIS}_{Y \chi p}$. We refer to a longer companion paper of ours in which analytical expressions of the relevant DIS cross sections are given~\cite{DeMarchi:2025uoo}.\footnote{
\label{foot:Wang2025}
As this work was being completed, Ref.~\cite{Wang:2025ztb} appeared on arXiv studying the mechanism that we first proposed in~\cite{DeMarchi:2024riu}, likewise for the case of vector mediator, but without addressing diffuse neutrinos. While~\cite{Wang:2025ztb} qualitatively confirmed our findings of~\cite{DeMarchi:2024riu} about DM-induced neutrino fluxes from TXS 0506+056, its results quantitatively differ from ours, for example the fluxes of~\cite{Wang:2025ztb} depend on $m_\DM$ at high $E_\nu$. Ours do not, as expected from our discussion in Sec.~\ref{sec:TXS}. See Sec.~\ref{sec:tests} and~\cite{DeMarchi:2025uoo} for more on~\cite{Wang:2025ztb}.}

\subsection{Diffuse astrophysical neutrinos from an ensemble of blazars}
\label{sec:diffuse_neutrinos} 
\begin{figure}[t]
    \centering
    \includegraphics[width=.5\textwidth]{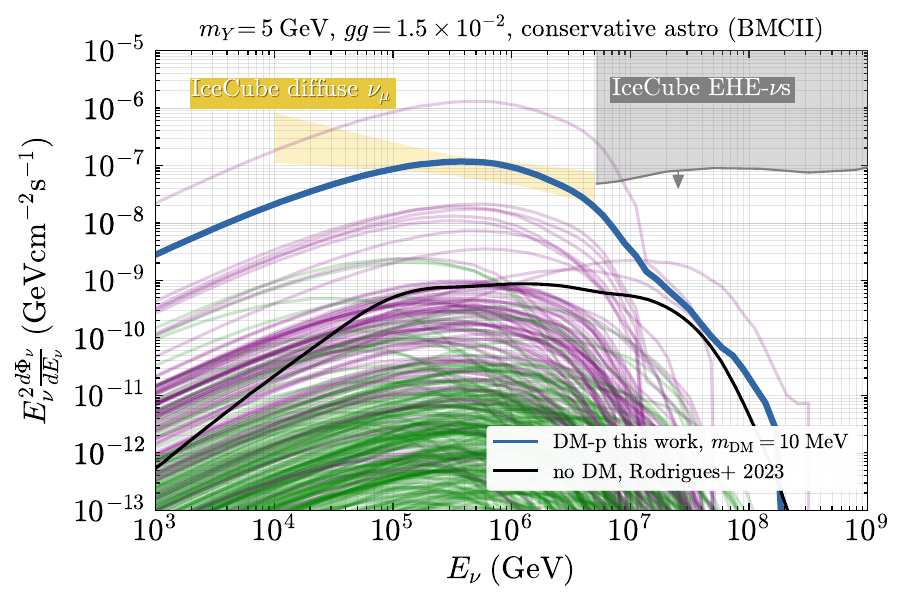}
    \caption{Neutrino flux from $p-\mathrm{DM}$ deep inelastic scatterings around 324 blazars that have been fitted with single-zone lepto-hadronic jet models in~\cite{Rodrigues:2023vbv}. The lines correspond to spin-1 DM mediators with mass $m_Y= 5 \;\mathrm{GeV}$, for $m_\DM = 10 \;\mathrm{MeV}$, $g_{\chi Y}g_{q Y} = 1.5\times 10^{-2}$, $R_\min = 10^4 R_S$.
    The case of spin-0 mediators is a simple rescaling of the spin-1, as in Fig.~\ref{fig:flux_comparison}. Blazars in the sample are separated into FSRQ (green) and BL Lac (purple) objects, the cumulative flux of all of them except the two outliers (see text) is shown as a thick blue line. In black is the neutrino flux computed in~\cite{Rodrigues:2023vbv} as a result of $p-\gamma$ interactions, with no DM. The gray shaded area is excluded by the null detection of extremely-high-energy (EHE) neutrinos in 9 year of IceCube data~\cite{IceCube:2018fhm}, the yellow band is as in Fig.~\ref{fig:flux_comparison}.}
    \label{fig:neutrino_flux_stacking}
\end{figure}

We then repeat the same procedure for each of the 324 blazars modelled in \cite{Rodrigues:2023vbv}, where the authors perform a lepto-hadronic fit to their multiwavelength electromagnetic emissions.
Our results for the DM-induced neutrino fluxes are displayed in Fig.~\ref{fig:neutrino_flux_stacking} for a common choice of DM parameters. The cumulative flux from 322 blazars is shown as a thick blue line.
The two purple lines that dominate below and above 10~PeV correspond to blazars 3C 371 and PKS B1413+135, respectively. We do not include them in the cumulative flux because their uniquely huge neutrino fluxes require a more careful and specific analysis than the global fit~\cite{Rodrigues:2023vbv}, which was carried out for other purposes. In addition, both blazars lie in the northern sky and so their PeV neutrinos, before reaching IceCube, suffer significant attenuation which is not included in our fluxes.

We find that the cumulative DM-induced neutrino flux can saturate the diffuse astrophysical neutrino flux observed by IceCube~\cite{Abbasi:2021qfz}, for DM parameter space allowed by all existing constraints (discussed in Sec.~\ref{sec:tests}).
We now comment on some crucial points and subtleties needed to fully appreciate the results of this section:
\begin{itemize}
    \item[$\diamond$] First and foremost, our results suggest the intriguing possibility that IceCube's diffuse astrophysical neutrinos above $O(100)$~TeV are explained by protons-DM interactions in blazars. 
    For that to be possible, one would need to find some correlation between IceCube's skymap and the one of blazars. 
    The analyses of \cite{Buson:2022fyf, Buson:2023irp} indeed do find a strong correlation in the southern sky in the energy range $E_\nu \gtrsim 100 \;\mathrm{TeV}$, and a smaller but still significant correlation for the northern sky and energies $E_\nu \lesssim 100 \;\mathrm{TeV}$, in accord with the expectations of our mechanism. However, the analysis of an updated dataset reported in~\cite{Bellenghi:2023yza} finds no significant correlation, although it is unclear to us whether this holds also for the highest-energy neutrinos predicted by our mechanism.
    The recent IceCube analysis~\cite{IceCube:2023htm} finds no significant correlation between high-energy neutrinos and a Fermi-LAT catalog of $O(2000)$ blazars, but it relies on the assumption that the neutrino flux can be described by a single power-law, which is not the case for our predicted fluxes. All in all, our proposal calls for implementing its specific characteristics in the searches for correlations between IceCube's skymap and the one of blazars. 
    \item[$\diamond$] 
    The DM-induced neutrino fluxes (purple and green lines in Fig.~\ref{fig:neutrino_flux_stacking}) have a significantly different shape than those for the same blazar sample but in the absence of DM (see Fig.~11 of~\cite{Rodrigues:2023vbv}, only their sum is shown in Fig.~\ref{fig:neutrino_flux_stacking}).
    This deserves some clarification.
    As in \cite{Rodrigues:2023vbv}, we separate blazars into their subcategories FSRQ (in green) and BL Lac objects (in purple). While \cite{Rodrigues:2023vbv} finds comparable fluxes from these two classes of objects, for us BL Lac objects completely dominate. The reason is that FSRQ have intense ambient photon fields for the protons to interact with and produce neutrinos, due to efficient accretion and bright broad line region (BLR), while BL Lac do not. As a consequence, to achieve a given neutrino flux in the absence of DM,
    FSRQ need a lower proton luminosity with respect to BL Lac, due to abundance of targets. For our mechanism, ambient photons are of no consequence and the only relevant parameter is the proton luminosity, thus breaking the degeneracy between number of protons and target photons in favour of BL Lac objects.
    \item[$\diamond$]
    Another difference between our neutrino fluxes and those of~\cite{Rodrigues:2023vbv} lies in the peak neutrino energy for each blazar. The photons produced by the BLR are in the eV-keV range, which requires  very large proton energies to reach the threshold for photo-meson production of neutrinos.
    Depending on how the photons are boosted/deboosted by the motion of the blob, the threshold moves to higher or lower proton energies, explaining the large variability in the neutrino peak energy. For our case, the threshold for neutrino production is always achieved for small proton energies because the DM has  a mass larger than the energies of BLR photons, thus neutrino production is always efficient. Furthermore, $\sigma^{\DIS}_{Y \chi p}$ increases with $\sqrt{s}$, thus the peak of the neutrino flux $E_\nu$ depends only on the maximal proton energy and the variability is greatly reduced. 
\end{itemize}

\begin{figure*}[p]
    \centering
        \includegraphics[width=.4\textwidth]{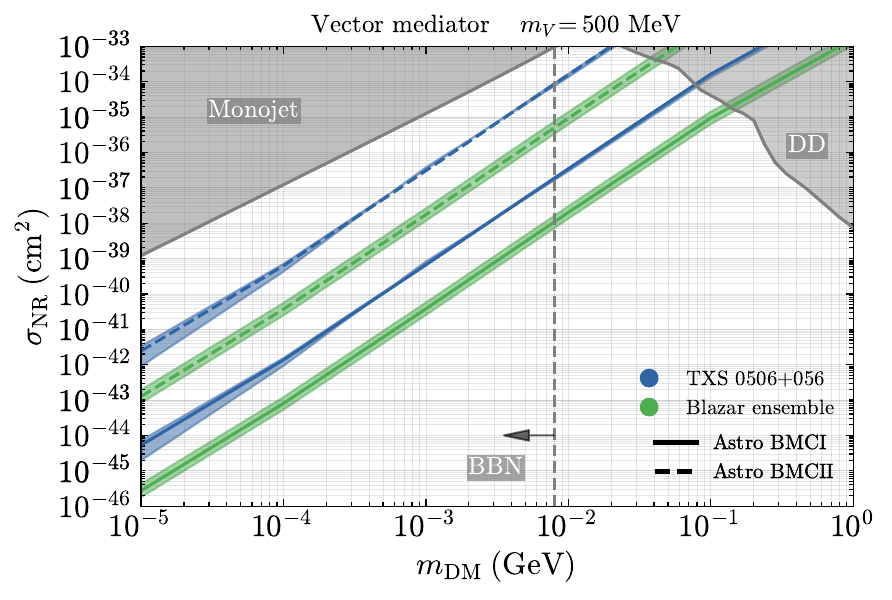} 
        \;
          \includegraphics[width=.41\textwidth]{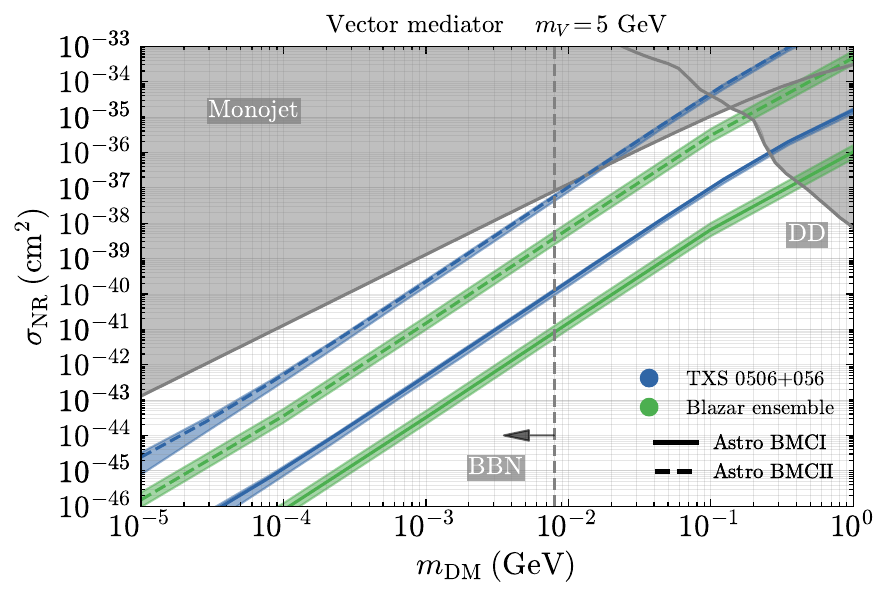} 
         \includegraphics[width=.41\textwidth]{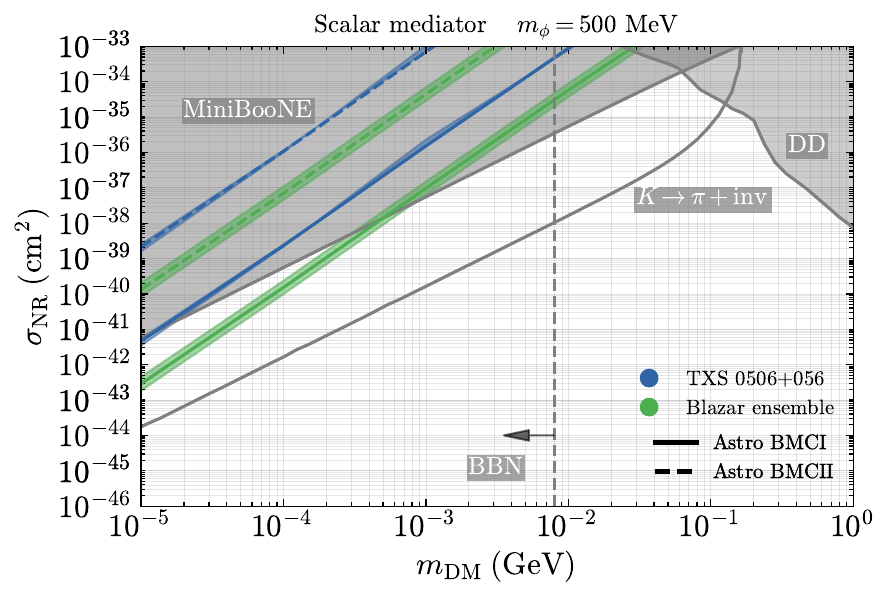}\;
         \includegraphics[width=.41\textwidth]{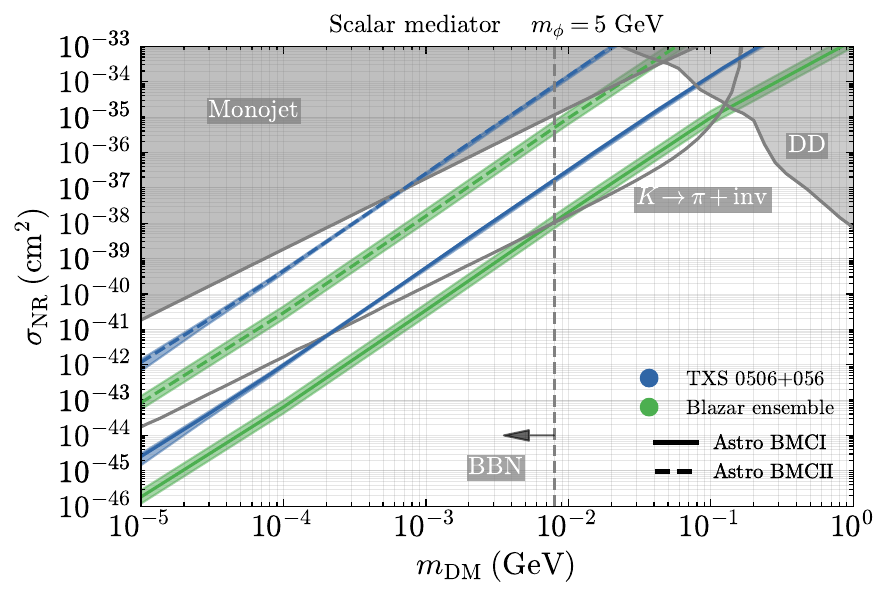}
        \includegraphics[width=.41\textwidth]{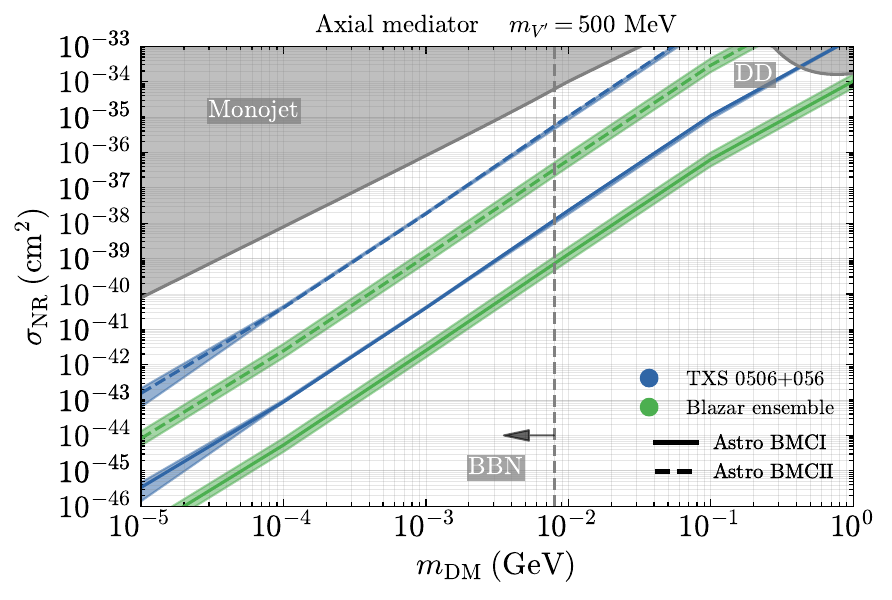} \;
        \includegraphics[width=.41\textwidth]{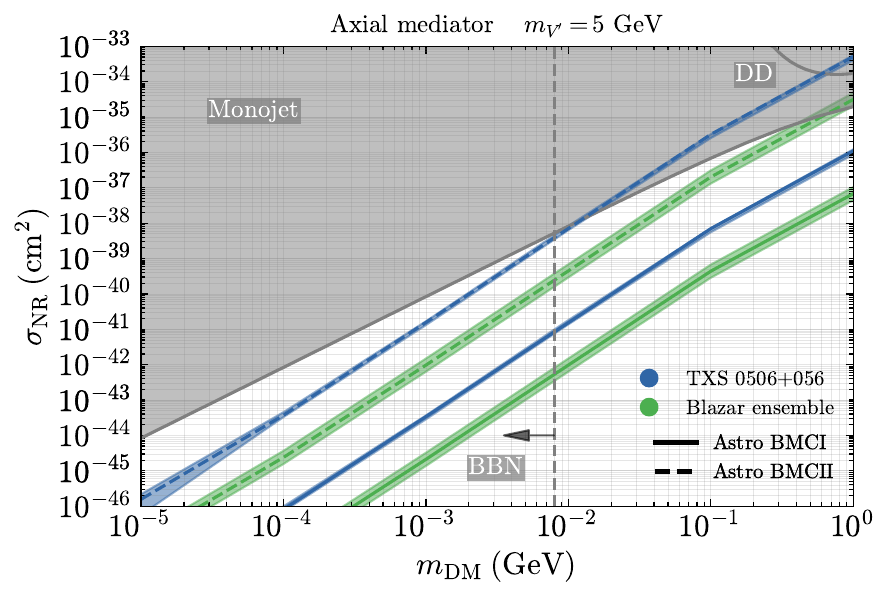} 
        \includegraphics[width=.41\textwidth]{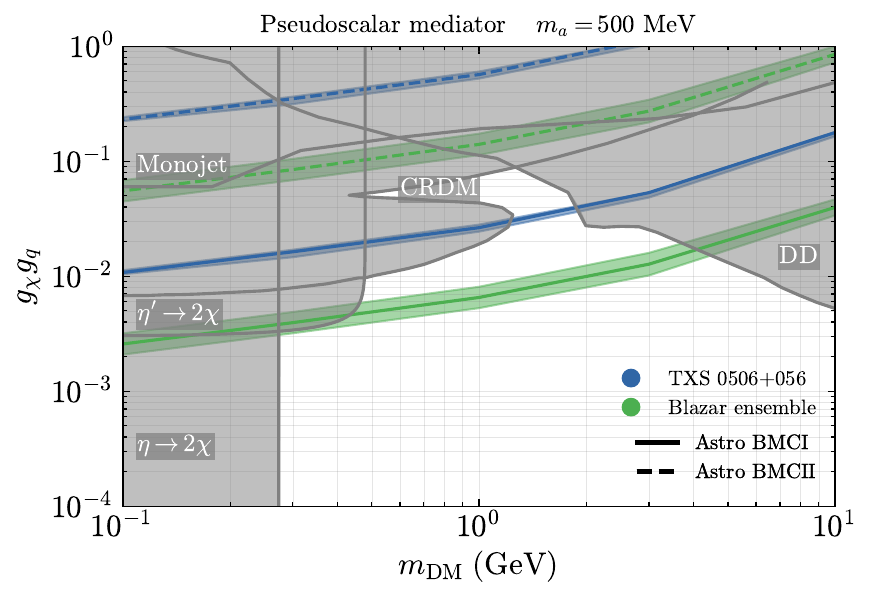}\;
        \includegraphics[width=.41\textwidth]{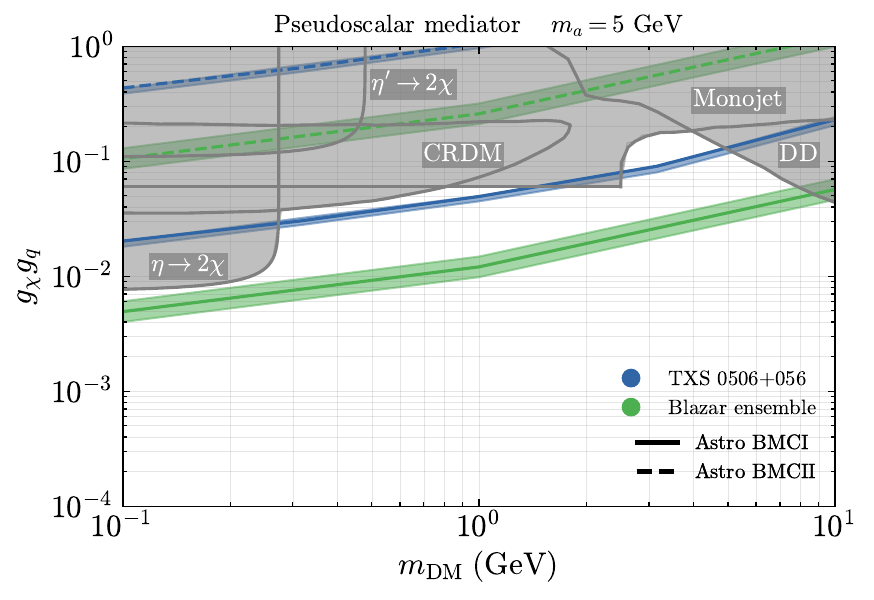}
    \caption{\footnotesize DM parameter space of     Eqs.~(\ref{eq:scalar})--(\ref{eq:axial}) for mediator mass $m_Y=500$~MeV (left column) and $m_Y=5$~GeV (right column), where $\sigma_\NR = (g_{\chi Y}^2 g_{N Y}^2/\pi)(\mu_{\chi N}^2/m_Y^4)$ and $g_{N Y}$ are functions of $g_{q Y}$ that we take from~\cite{DeMarchi:2025uoo}.
    High-energy diffuse neutrinos~\cite{Abbasi:2021qfz} and the IC-170922A neutrino from TXS 0506+056~\cite{IceCube:2018dnn} are explained, respectively, along the green and blue lines, where their width accounts for uncertainty on the energy $E_\nu$ of the TXS 0506+056 neutrino and on the diffuse neutrino flux, which itself is shown as a shaded band in Fig.\ref{fig:flux_comparison}, \ref{fig:neutrino_flux_stacking}. Continuous (dashed) lines correspond to BMCI (BMCII) for the DM LOS integral, see Sec.~\ref{sec:LOS}.
    The parameter space on the left of the `BBN' dashed line is excluded by big bang nucleosynthesis~\cite{Sabti:2019mhn}, unless the universe is reheated below the QCD scale~\cite{Berlin:2018ztp}.
    Direct detection (DD) of halo DM~\cite{SuperCDMS:2023sql,CRESST:2019jnq,DarkSide:2018bpj,LZ:2024zvo,NEWS-G:2024jms,XENON:2023cxc, XENON:2025vwd} excludes the gray regions in the top-right of each plot. Note that DM-nucleon scatterings are spin-independent (SI) if induced by $V$ and $\phi$, spin-dependent if by $V'$, and momentum and spin-dependent if by $a$, so that DD limits in the $a$ case come from loop-induced SI scatterings~\cite{Abe:2018emu}.
    The other gray regions are excluded as follows. $V$: by monojet at Tevatron and LHC \cite{Shoemaker:2011vi}; $\phi$: by DM searches from proton beam dump at MiniBooNE \cite{MiniBooNE:2017nqe} recast for a scalar mediator decaying invisibly \cite{Batell:2018fqo}, by monojet at LHC~\cite{ATLAS:2020wzf} recast in~\cite{Ema:2020ulo} and by $K\to\pi$ invisible decays~\cite{Cox:2024rew} which we do not shade as it can be evaded by specific combinations of $g_{u\phi}$ and $g_{d\phi}$~\cite{Pascoli:atmospheric}; $V'$: by monojet at Tevatron and LHC~\cite{Shoemaker:2011vi}; $a$: by monojet at LHC~\cite{ATLAS:2020wzf,Ema:2020ulo}, invisible meson decays~\cite{ParticleDataGroup:2024cfk} and CR upscattering (CRDM)~\cite{Ema:2020ulo}.
    All lines and limits depend on $g_{\chi Y} g_{q Y}$, except monojet limits if $m_\DM < m_Y$, in which case we assume $g_{\chi Y} = 1$.
    See Sec.~\ref{sec:tests} for more details.}
    \label{fig:Constraints}
\end{figure*}

\section{Tests of dark matter origin for $\nu$'s}
\label{sec:tests}

We find that DM-$p$ inelastic scatterings around blazars could explain IceCube's observations, of both diffuse neutrinos and of the neutrino from TXS 0506+056, for several choices of DM models and parameters. These are indicated in Fig.~\ref{fig:Constraints} by four lines per plot, where each line corresponds to either TXS 0506+056 or the diffuse neutrinos and to one of our two benchmarks for $\Sigma_{\DM}^\text{spike}$ of Sec.~\ref{sec:LOS}. Each plot is then repeated for the four mediators of DM-quark interactions (see Sec.~\ref{sec:toymodels}) and for two values of the mediator mass.

While the lines explaining TXS 0506+056 do not overlap with those explaining the IceCube diffuse neutrinos, it would be incorrect to conclude that the two observations cannot be explained at once for the same DM particle model and parameters. Indeed, these lines could be easily brought to overlap by minor modifications in the jet parameters of some of the blazars involved in this study, a possibility which would be interesting to explore in future fits of blazar jets that include our DM mechanism.
Independently of jet parameters, the lines could also overlap because of minor modifications in the line-of-sight integral of DM $\Sigma_{\DM}^\text{spike}$ around specific blazars, because that integral depends on blazar-specific properties like their recent history and their exact starting jet points, and our choice of a single $\Sigma_{\DM}^\text{spike}$ value for all of them has been driven by simplicity.

In Fig.~\ref{fig:Constraints} we also display, to the best of our knowledge, existing limits on the same DM models from cosmology, from DM-induced recoils and from searches of DM produced in the lab. 
These limits are described in short in the caption, we refer the reader to our companion work~\cite{DeMarchi:2025uoo} for more details on them. 
Our comprehensive exercise teaches us that DM could be the origin of (both TXS 0506+056 and diffuse) IceCube's neutrinos for several different DM dynamics, proving the generality of our proposal.
It also informs us that missing-energy searches in the laboratory are a generic test of our mechanism. Other tests depend instead on the DM dynamics under consideration. For example, the vector- and scalar-mediated models will be best tested (each one to different extents) by direct searches for halo DM, while the pseudoscalar-mediated one by searches for CR-upscattered DM like~\cite{Ema:2020ulo,Super-Kamiokande:2022ncz}. 
These will also be important in the case DM is inelastic (see e.g.~\cite{CarrilloGonzalez:2021lxm,Berlin:2023qco} for recent studies), where our proposal still works, but that we have not explicitly included in our toy models to keep the discussion contained. Indeed, inelastic DM evades halo-DM detection but not searches for DM signals that rely on larger exchanged momenta, like our DM-induced neutrinos here, CR-upscattering and accelerator probes.

Other DM-induced signals from the same blazars depend on exactly the same DM and astrophysical parameters as the neutrino signals discussed in this work, and provide therefore a robust test of our mechanism.
A first such example is given by blazar-boosted DM, which can be detectable at direct detection facilities and neutrino detectors, as first shown in~\cite{Wang:2021jic, Granelli:2022ysi}. We leave its study for a companion paper~\cite{DeMarchi:2025uoo} because of the large amount of different physics involved (both in DM upscattering and in its signals at neutrino detectors) and because its study also answers questions that are independent of our mechanism to explain IceCube's neutrinos.
A second example of signals induced by the same DM-$p$ scatterings in blazars is given by the photons that they produce, see~\cite{Gorchtein:2010xa} for a pioneering study. At the source and before any propagation, these photons are indeed comparable to the neutrinos computed here, because they also mostly come from the pions produced in the hadronic showers. The determination of such photon fluxes on Earth is a necessary future step to test our mechanism, but it requires to implement exactly the same assumptions (on photon propagation, in particular on the astrophysical environment close to, and far from, the blazar) as the blazar fits that we employed, otherwise it would not be consistent.
This exercise then goes far beyond the purposes of this paper, and we leave it for future work.
\footnote{Note that it is not consistent either to compute the photon spectrum from DM-$p$ DIS in jets and include, in its propagation to the Earth, only the effects from CMB and extragalactic background light, as done in~\cite{Wang:2025ztb} (see also footnote~\ref{foot:Wang2025}).
This indeed ignores interactions in the environment close to the blazar which, from studies of blazar jets (see e.g.~\cite{Rodrigues:2018tku,Cerruti:2018tmc}), are known to importantly affect photon spectra on Earth.}

\section{Conclusions}
\label{sec:summary}

The origin of high-energy neutrinos, both diffuse astrophysical ones and those associated with blazar, is not understood. The widely adopted one-zone lepto-hadronic models of blazar jets, for example, fall short in reproducing both observations.

In this paper we have proven that deep inelastic scatterings, between protons within the same jets and DM around blazars, can be the origin of both the diffuse neutrino flux for energies above 100 TeV and of the neutrino observed from TXS 0506+056. We have found that this mechanism is compatible with existing DM searches for four different mediators of DM-quark interactions, proving its generality and identifying future tests of each DM dynamics.
While our proposed DM explanation of diffuse neutrinos is, to our knowledge, a novel result, the one of the neutrino from TXS 0506+056 constitutes an extension of our~\cite{DeMarchi:2024riu} to a larger set of DM-quark interactions.

An alternative explanation of the observed high-energy neutrinos, at least those from TXS 0506+056, consists in multi-zone models for blazar jets~\cite{Liu:2018utd, Righi:2018xjr, Sahakyan:2018voh, Wang:2018zln, Zhang:2019dob, Xue:2019txw, Zhang:2019htg, Wang:2024tsd}. They involve external structures outside the blobs and thus introduce more parameters than the single-zone models employed in our study.
The DM scenario studied here also introduces more parameters, but it can be independently tested by searching for DM interactions taking place under controlled conditions on Earth. This underlies the importance of the analysis of different DM-$p$ interactions, carried out in this paper.

Our findings open several avenues of investigation. On an observational side, they motivate the search for correlations between high-energy neutrinos and the blazars that dominate our flux. On an astrophysical side, they call for a calculation of the photon spectrum induced by the same DM-proton interactions, which we leave for future work. This would allow to determine possible changes in the best-fit parameters of blazar jets, in particular their proton luminosities, that without DM typically exceed the Eddington limit $L_{\rm Edd} \approx 1.26\times 10^{38} (M_{\BH}/M_{\odot}) \,\text{erg/s}$ (i.e. the maximal power compatible with hydrostatic equilibrium of accreting matter, see e.g. \cite{Dermer:2009zz}).
Finally on a particle-physics side, our results motivate searches for blazar-boosted DM fluxes, that we assess in the companion paper~\cite{DeMarchi:2025uoo}, and more in general for sub-GeV DM.

\section*{Acknowledgements}
This work was supported in part by the European Union's Horizon research and innovation programme under the Marie Skłodowska-Curie grant agreements No.~860881-HIDDeN and No.~101086085-ASYMMETRY, by COST (European Cooperation in Science and Technology) via the COST Action COSMIC WISPers CA21106, and by the Italian INFN program on Theoretical Astroparticle Physics. FS thanks the Galileo Galilei Institute for theoretical physics (GGI) for kind hospitality during the completion of this work. AGDM is grateful to the Lawrence Berkeley National Laboratory for its kind hospitality during the completion of this work.

\bibliography{Blazar_bib}
\bibliographystyle{JHEP}

\end{document}